# Far-field characterization of the thermal dynamics in lasing microspheres

J. M. Ramírez[1,a)], D. Navarro-Urrios[2,a)], N.E. Capuj[3], Y. Berencén[1], A. Pitanti[2], B. Garrido[1], A. Tredicucci[4]

[1] MIND-IN2UB, Dept. Electrònica, Universitat de Barcelona, Martí i Franquès 1, 08028, Barcelona, Spain

[2] NEST, Istituto Nanoscienze – CNR and Scuola Normale Superiore, Piazza San Silvestro 12, Pisa, I-56127, Italy

[3]Depto. Física, Universidad de la Laguna, 38206, Spain

[4] NEST, Istituto Nanoscienze and Dipartimento di Fisica, Università di Pisa, Largo Pontecorvo 3, I-56127 Pisa, Italy

a) These authors contributed equally to this work.

*This work reports the dynamical thermal behavior of lasing microspheres placed on a dielectric substrate while they are homogeneously heated-up by the top-pump laser used to excite the active medium. The lasing modes are collected in the far-field and their temporal spectral traces show characteristic lifetimes of about 2 ms. The latter values scale with the microsphere radius and are independent of the pump power in the studied range. Finite-Element Method simulations reproduce the experimental results, revealing that the thermal dynamics is dominated by the heat dissipated towards the substrate through the medium surrounding the contact point. The characteristic system scale regarding thermal transport is of few hundreds of nanometers, thus enabling an effective toy model for investigating heat conduction in non-continuum gaseous media and near-field radiative energy transfer.*

Whispering gallery mode (WGM)-based lasing cavities can enhance the high sensitivity of passive WGM cavities [1,2] and achieve extremely low detection limits thanks to their narrow linewidths [3-5]. Input bus-waveguides for injecting light inside the cavity are not necessary and simple far-field pumping configurations can be used. Regarding signal collection, the high intensity of the emitted laser light allows for either standard far-field detection or for schemes using output coupled bus-waveguides [6,7]. Made of semiconducting materials, high-quality WGM cavities are prone to thermally-induced changes due to the thermo-optic (TO) effects, i.e., thermo-refractive mechanisms and thermal expansion, which often limit their sensitivity [5]. Moreover, a wide range of thermally induced nonlinear phenomena such as hysteretic wavelength response and oscillatory instability have been reported experimentally in such cavities [8-10].

Although usually considered detrimental features, TO effects can be exploited in time-resolved experiments for extracting heat exchange rates between the cavity and the external environment [10-12]. In spite of the lack of large sensitivities, these techniques allow extracting quantitative information of the medium surrounding the cavity without the need of calibrating the resonance spectral shift in relation to the changing magnitude. Heat transfer dynamics using WGM passive cavities has been successfully studied to measure adsorbed water layer thickness and desorption rates [11] and to obtain absolute values of thermal accommodation coefficients of different gases at low pressures [12]. In particular, a wide interest has been reported in predicting heat transfer to a surface when immersed in a gas in the non-continuum limit, when the characteristic length scale of the system is comparable to the gas mean free path. This limit can be used to describe the gas environment in a variety of applications, ranging from MEMS devices (low dimensions) to semiconductor manufacturing or spacecraft aerodynamics (low pressures) [13].

Standard investigation of dynamical thermal behavior of passive WGM cavities is performed by in-coupling laser light through the evanescent field of a tapered fiber. Under such conditions, the transmitted light is monitored while the WGM cavity is out of thermal equilibrium in response to non-adiabatic heating [11,12]. Conversely, the coupled light can be used both to test the thermal dynamics and as a heating source localized within the WGM volume [10]. However, this approach highly complicates the dynamics of the system since: i) the heating efficiency is modulated by the shape of the resonance and ii) the cavity takes some time to thermalize internally.

In this work, we present an alternative approach that combines high-power laser emission and narrow spectral linewidths with excitation-detection in the far field, enabling a straightforward investigation of the thermal dynamics of the cavity with all-optical means.

The investigated WGM lasing cavity is a Nd-doped Barium-Titanium-Silicate (BTS) [7] microsphere fabricated with the method reported in ref. [14] placed on top of a 2.7 micron thick $SiO_2$ layer deposited over a Si substrate. A free space laser is used to resonantly excite the $Nd^{3+}$ ions and homogeneously heat the cavity, while the signal of a lasing WGM is collected in the far-field to monitor the thermal dynamics of the cavity. The sphere-plate geometry has been the object of deep investigation during the last years in the context of local thermal transfer between non-planar geometries [15-17] mainly driven by their promising application in nanoscale heat stamps [18-19]. We demonstrate that, at ambient conditions, the thermal dynamics is governed by heat conduction to the substrate, mediated by the medium surrounding the contact point. Given that the characteristic dimensions of thermal transport are of few hundred nanometers, i.e., substantially lower than the thermal wavelength and comparable to the gas mean free path, we propose lasing microspheres as a simple and versatile scheme to investigate radiative thermal conduction and non-continuum gas thermal behaviors.

**Results and discussion**

The measurement setup for the optical experiment is schematically shown in Figure 1. A continuous wave (CW) laser at 808 nm was focused on a single microsphere (radius between 20 and 25 µm) by means of a microscope objective [20], the spot size being large enough to guarantee homogeneous spatial pumping over the microsphere volume. The PL signal associated to WGMs contained in a plane parallel to the substrate was collected by a second microscope objective, focused on the input slit of a monochromator and recorded either by a charge-coupled-device (CCD) detector or a photomultiplier tube (PMT). Time-resolved experiments were made by modulating the pump with a mechanical chopper and by registering the PMT output with an oscilloscope.

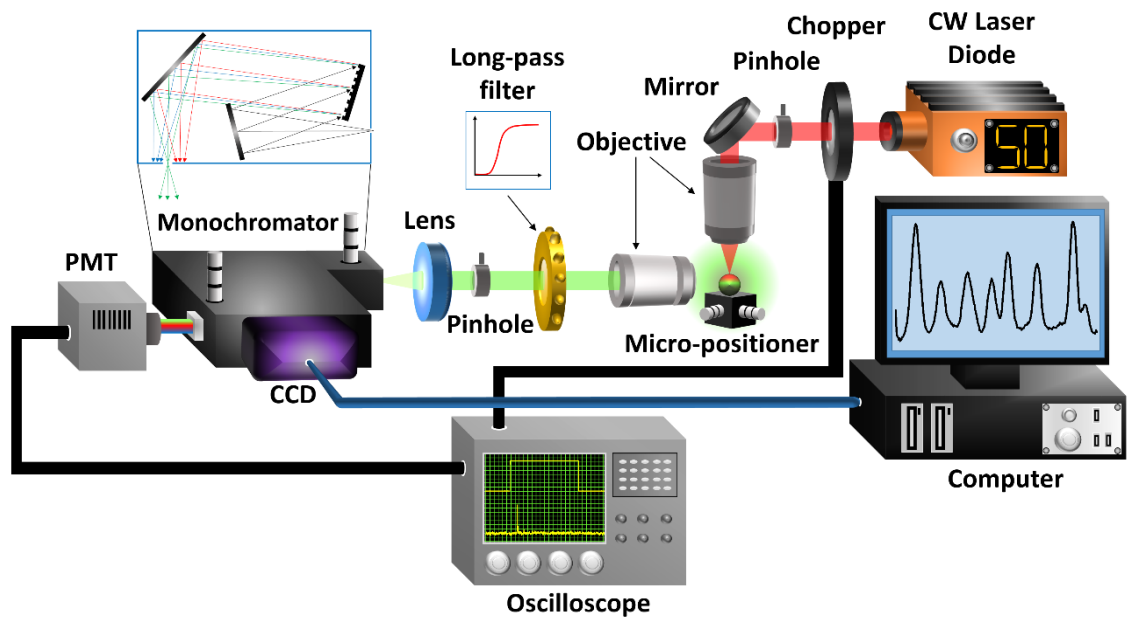

**Figure 1.** Schematic diagram of the experimental set-up. Dimensions are not to scale.

The $Nd^{3+}$ excitation process is described in Figure 2a. Under external excitation, $Nd^{3+}$ ions are promoted from the ground state ($^4I_{9/2}$) up to the $^4F_{5/2}$ and $^2H_{9/2}$ levels, from where a rapid non-radiative thermalization to the $^4F_{3/2}$ transition takes place. Subsequently, another radiative de-excitation occurs from this level down to the $^4I_{11/2}$ state. The final thermalization to the ground state is through non radiative scattering. $Nd^{3+}$ effective dynamics can be modeled by a four-level rate equation system, where $N_4$, $N_3$, $N_2$ and $N_1$ correspond to the number of ions excited in transitions $^4F_{5/2}$, $^4F_{3/2}$, $^4I_{11/2}$ and $^4I_{9/2}$, respectively. Laser action at around 1064 nm occurs if $N_3 > N_2$ and the passive losses are compensated.

Figure 2b shows a 2D color-map where the measured PL intensity of a microsphere with radius R=25 µm is displayed as a function of the observation wavelength ($\lambda_c$, y-axis scale) and the pump photon flux ($\Phi_p$, x-axis scale). Multimode lasing is observed at various wavelengths within the $Nd^{3+}$ emission band at around 1064 nm. A typical spectrum taken at the maximum used photon flux ($\Phi_p$ = 5.5x10$^{23}$ ph/cm$^2$s) is reported in the right hand side of Figure 2b. The two most intense lasing peaks have been numbered using 1 for the main peak.

In the same figure, a red-shift associated to the TO effect is observed for all lasing modes when $\Phi_p$ is increased. In a first order approximation it is possible to relate the spectral shift of the lasing mode ($\Delta\lambda = \lambda_s - \lambda_{ini}$, where $\lambda_s$ and $\lambda_{ini}$ are the stationary lasing wavelength for a given temperature and the cold-cavity resonance wavelength respectively) with the temperature difference volume where the mode is localized and the surrounding environment as:

$$\Delta\lambda = a\Delta T \qquad (1)$$

where $a$ is the TO coefficient, accounting both for physical dilatation and refractive index changes [10]. Figure 2c shows, for a microsphere with radius R=25 µm, that while laser emission is inherently nonlinear (at first superlinear and then saturated for high $\Phi_p$ (left y-axis scale)), $\Delta\lambda$ follows $\Phi_p$ linearly (right y-axis scale). This indicates that the dominant heating mechanism is partial absorption of the incident pump power radiation ($P_{abs} = \sigma_p \Phi_p N_1 V \frac{hc}{\lambda_p}$, where $\sigma_p$, $\lambda_p$ and $V$ are the absorption cross section, pump wavelength and sphere volume respectively), which is to some extent released into heat through the non-radiative decay of the excited ions described above. This result lets us to neglect localized heating within the cavity mode and therefore to consider spatially homogeneous heating of the whole sphere. The offset of the linear spectral shift curve corresponds to the cold-cavity resonance ($\lambda_{ini}$), which is $\lambda_{ini}$ = 1065.0 nm for the measured sphere.

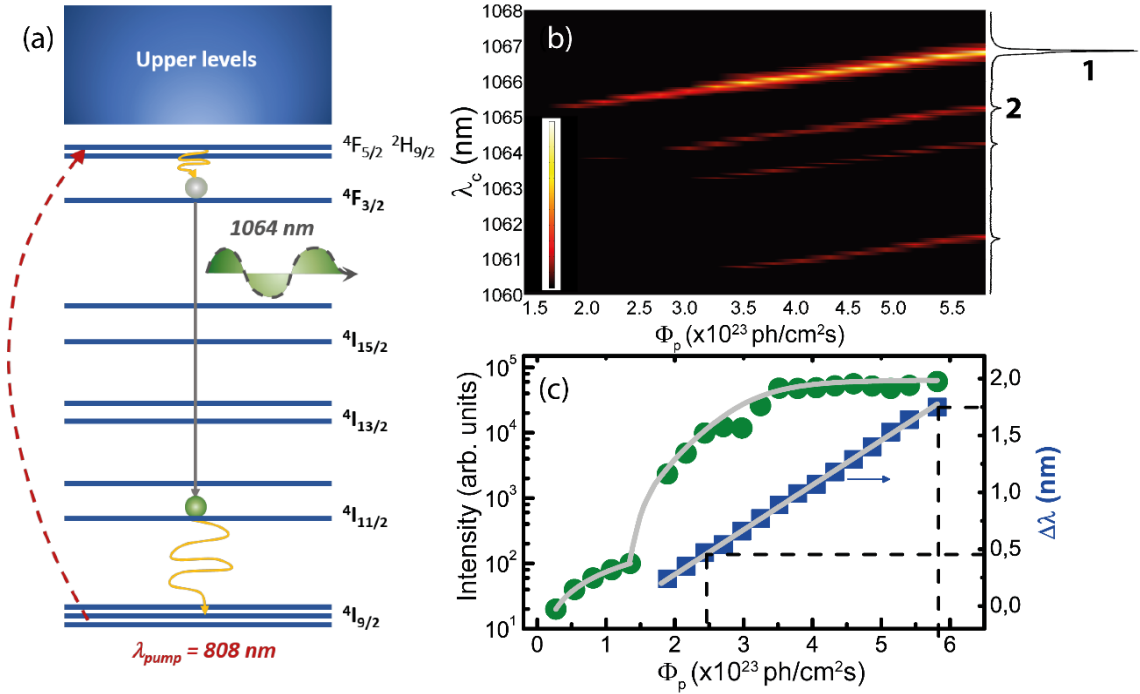

**Figure 2. (a)** Schematic diagram of $Nd^{3+}$ energy levels under excitation at 808 nm. **(b)** 2D color-map of the PL intensity (in color) as a function of the wavelength (y-axis) and the pump photon flux (x-axis). **(c)** Intensity of a lasing resonance (left y-axis scale) and wavelength shift relative to the value of the cold-cavity resonance wavelength (right y-axis scale) as a function of the incident photon flux. Dashed lines mark the values of the incident photon flux at which time-resolved measurements were performed (see Figure 4).

The thermal dynamics of the microsphere can be described by the Fourier heat equation:

$$C_p \frac{d(\Delta T)}{dt} = \gamma P_{abs} - k\Delta T , \qquad (2)$$

where $C_p$ is the heat capacity of the microsphere ($C_p=\rho V c_p$, $\rho$ and $c_p$ being the mass density and the specific heat of the BTS glass respectively), $\gamma$ is the efficiency at which $P_{abs}$ is converted into heat and $k$ is the effective thermal conductivity between the cavity mode volume and the surrounding. In order to extract an analytic solution to the previous equation we assume that the dynamics governing the ion level population has achieved a stationary regime, making the heat absorbed rate time-independent. Under such condition, and using Eq. 1, the temporal evolution of $\Delta\lambda$ can be written as follows:

$$\Delta\lambda(\Delta t) = a\frac{\gamma P_{abs}}{k}\left(1 - e^{-\frac{k}{C_p}\Delta t}\right) \qquad (3)$$

The exponential fit of the dynamical $\Delta\lambda$ evolution allows extracting the transitory characteristic time ($\tau = \frac{C_p}{k}$). This scales linearly with $R$, since $C_p$ and $k$ are proportional to the sphere volume and surface respectively if the heat leaking through the contact point is neglected. The stationary spectral shift is given by:

$$\Delta\lambda(\Delta t \to \infty) = \lambda_s - \lambda_{ini} = a\frac{\gamma P_{abs}}{k}. \qquad (4)$$

The experimental linear evolution with $\Phi_p$ agrees with the theoretical analysis, which supports the hypothesis of considering $k$ as a temperature-independent parameter. This is confirmed by the large stability of thermal conductivity of air at the experimental conditions (50% relative humidity), with a 10% variation over hundred degrees above ambient temperature [24].

Time-resolved measurements of the laser emission spectral shift were carried out in phase with the modulated pump. Different $\lambda_c$ can be collected, accordingly to the monochromator filtering. A transitory signal is detected when the lasing resonance aligns with $\lambda_c$ on its pathway from $\lambda_{ini}$ towards $\lambda_s$ as a consequence of the thermal-induced red-shift. The red-shift delay time (Δt) can be extracted taking the chopper as an absolute reference marking the beginning of the microsphere excitation process. A complete temporal evolution of the lasing wavelength spectral shift can be obtained by subsequent measurements of Δt for a $\lambda_c$ sweep of the full spectral trajectory in small steps (0.01 nm). Figure 3 explains the followed methodology by illustrating three different situations:

- In panel a) ($\lambda_c \approx \lambda_{ini}$) the time-resolved signal (red solid line on the low part of the panel) shows a sharp signal overshoot that is detected when the laser pump switches on.
- Panel b) ($\lambda_{ini} < \lambda_c < \lambda_s$) shows an intermediate situation in which an asymmetric overshoot signal is captured. This is ascribed to a decrease of the red-shift rate during the measurement, i.e., the resonance enters into the detection range at a higher rate than when it comes out.
- Panel c) ($\lambda_c \approx \lambda_s$) shows that a stationary signal appears and persists for the entire excitation period until the pumping is switched off again. This corresponds to a full thermalization of the microsphere, the stationary lasing wavelength being that measured with CW excitation.

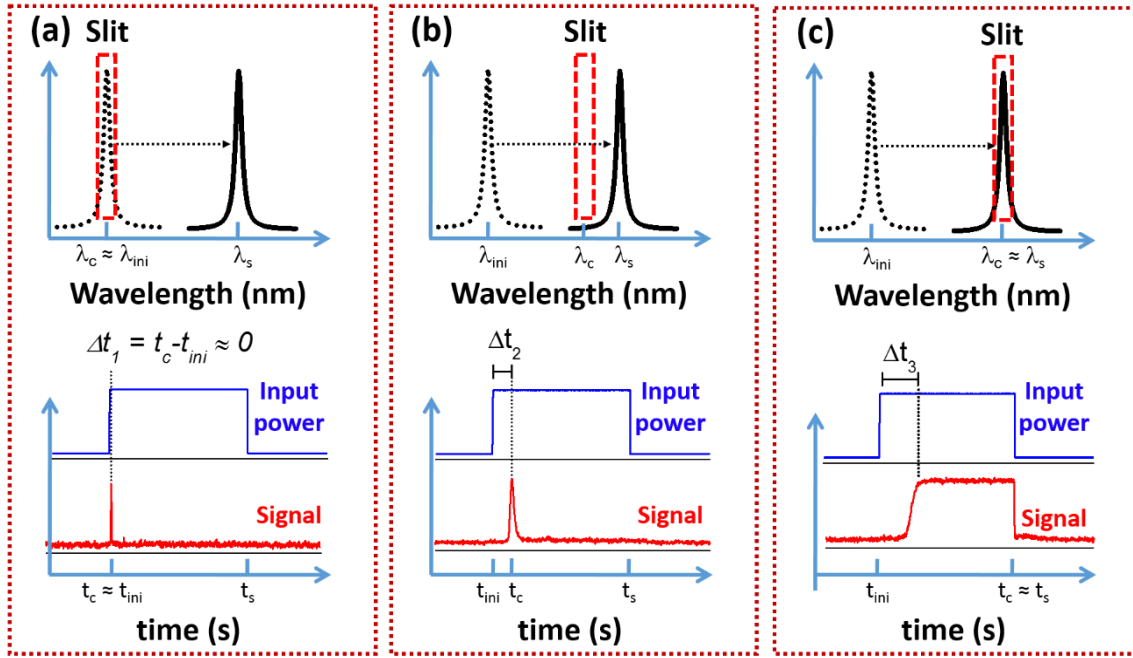

**Figure 3.** (Top panels in a, b and c) Schematic diagram of the relative position between collection wavelength (red dashed rectangle) and lasing wavelength shift range. Dotted resonances depict the cold-cavity resonance wavelength and solid resonances the stationary wavelength for a given photon flux. Bottom panels in a, b and c detail the observed line-shape of the time-resolved signal under three different wavelengths of detection: **(a)** at $\lambda_c \approx \lambda_{ini}$, **(b)** at $\lambda_{ini} < \lambda_c < \lambda_s$ and **(c)** at $\lambda_c \approx \lambda_s$. Notice that $\Delta t_1 < \Delta t_2 < \Delta t_3$.

The 2D color-map plot of Figure 4a illustrates a particular case obtained by pumping a sphere of R=25 μm with $\Phi_p$ =5.5×10$^{23}$ ph/cm²s. The spectrum covers the full temporal trace of lasing mode 1 and the last part of that of lasing mode 2 (see the corresponding CW spectrum in Figure 1b). Both curves follow the same exponential law, reaching $\lambda_s$ after about 10 ms. From this point on, stable thermal equilibrium is maintained.

Figure 4b shows the temporal traces of the spectral position of lasing mode 1 upon low and high $\Phi_p$ (2.5×10$^{23}$ ph/cm²s and 5.5×10$^{23}$ ph/cm²s, respectively). As expected, $\Delta\lambda_{max}$ scales with $\Phi_p$, in accordance with Eq. 4 and with the previously reported values in Figure 2c (marked with dashed lines). When normalizing $\Delta\lambda$ to $\Delta\lambda_{max}$ for the specific value of $\Phi_p$, the obtained curves overlap (Figure 4c). In fact, a characteristic lifetime of $\tau$ (R=25 μm) = 2.4 ms is extracted for both curves, which, according to Eq. 4, allows us to conclude that also $C_p$ can be assumed independent of the temperature of the cavity. Indeed, it has been previously demonstrated that although $\Delta T$ can reach a few hundreds of degrees in extreme cases [21], $C_p$ in BTS is almost unresponsive to temperature variations [22].

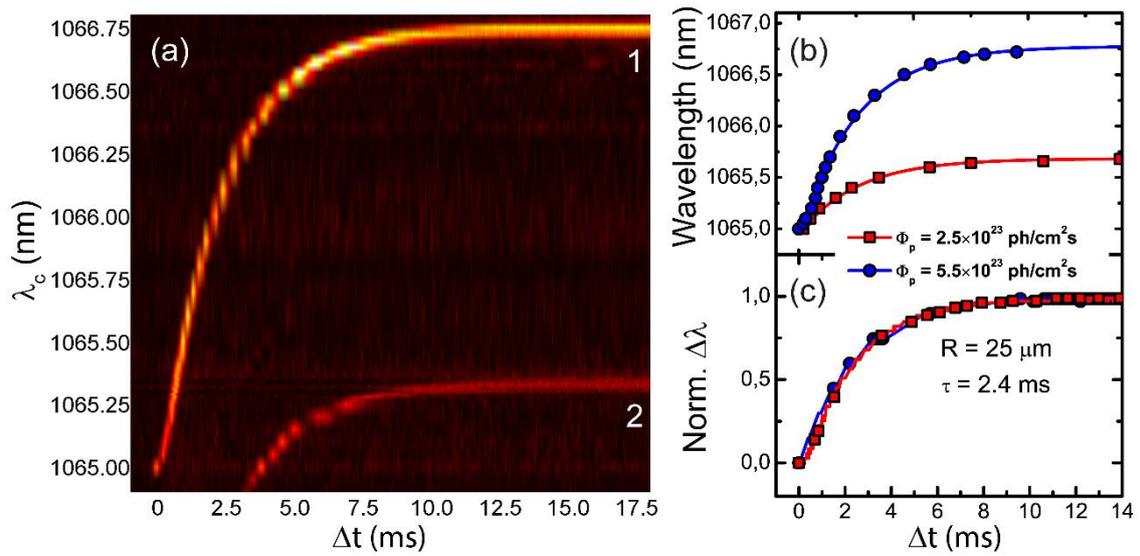

**Figure 4. (a)** Time-resolved 2D color-map of the wavelength red-shift of lasing mode 1 (and part of mode 2) under an incident photon flux of $5.5\times 10^{23}$ ph/cm²s. Colors refer to detected PL intensity, being the legend the same as in Figure 1b, y-axis scale represents the collection wavelength and the x-axis scale is the excitation time, its origin taken at $t_{ini}$. **(b)** Temporal behavior of the laser spectral position of mode 1 for low ($2.5\times 10^{23}$ ph/cm²s) and high ($5.5\times 10^{23}$ ph/cm²s) incident photon fluxes, respectively. **(c)** Normalized wavelength shift for low and high photon fluxes together with a fit using Eq. 3.

Finally, in Figure 5a we have compared the thermal dynamics of a sphere of R=25 μm ($\tau$(R=25 μm) = 2.4 ms, black dots), with that of a smaller radius (R=20 μm, green dots). As expected, the characteristic lifetime scales with R, the value obtained for the latter case being $\tau$(R=20 μm) =1.8 ms. It is worth noting that an equivalent trend was observed in several lasing spheres of similar radii, denoting high consistency and reproducibility.

In order to get more insight on the origin of the observed thermal dynamics we have performed Finite-Element Method (FEM) simulations on a system consisting of a BTS glass microsphere that is heated up by a homogenous source distributed along the sphere volume. The used FEM module solves Fourier's law of heat conduction just considering conductive heat transfer mechanisms through purely diffusive processes. We have neglected convection on the basis of the small dimensions of the spheres [23] and radiation since it only becomes significant with respect to conduction at gas pressures below $10^{-2}$ mb [17]. On the basis of the high relative humidity of the ambient, we have considered a temperature independent air thermal conductivity, as discussed previously.

We have first solved the problem of a sphere fully surrounded by air (orange solid curve of Figure 5a). The obtained result is a spatially homogenous output heat flux along the sphere surface but a much slower thermal dynamics than that observed experimentally. The second case of study was that of a sphere under vacuum deposited on a substrate such as that of the experiment, which leads to an even slower dynamics (cyan solid curve of Figure 5a). This is an expected result, as the released heat can only be dissipated through the contact point between the sphere and the substrate. The direct combination of both cases, i.e., microsphere surrounded by air on top of a substrate, successfully reproduces the experimental results for the two microsphere radii under analysis ($R_1$ = 20 μm and $R_2$ = 25 μm). It is worth noting the absence of free parameters, the input values being only the geometric dimensions and the intrinsic properties of the materials involved. Figure 5b shows the corresponding stationary $\Delta T$ profiles along the horizontal (r, parallel to the substrate surface) and vertical (z, perpendicular to the substrate surface) directions passing through the center of the sphere , i.e. r=0 and z=0. The inset shows the bidimensional distribution along a cross section of the whole system geometry. In the horizontal direction (along the r-axis) the temperature profile (black curve of Figure 5b) is isotropic. Starting from r = 0, a subtle decrease in $\Delta T$ is observed when approaching the surface of the sphere, followed by a steep decay for r values beyond the radius of the sphere (r > 25 µm in this case). On the contrary, along the vertical direction, the profile is strongly affected by the substrate, providing a highly asymmetric feature. In the upper semisphere, $\Delta T$ is much higher while it gets very close to zero at the contact point with the substrate. This spatial distribution is better understood by analyzing the output heat flux across the microsphere surface (Figure 5c), which reveals that most of the heat dissipates through its lower part. It is thus possible to conclude that the main heat loss mechanism is thermal conduction towards the substrate through the region surrounding the substrate contact point. In fact, temperature gradients increase towards it, given that the substrate remains very close to ambient temperature. It is worth noting that the temperature variation within the region around the contact point is much lower than the average temperature of the lasing mode, since the latter is localized in the equator of the microsphere (Figure 5b). The characteristic length scale of the system, which is the distance between the lower part of the sphere surface around the contact point and the substrate ($\Delta z=R(1-\cos\theta)$, where $\theta$ is the angle with respect to the z axis pointing downwards), remains below 400 nm in the region dissipating 90% of the heat ($\theta<11°$).

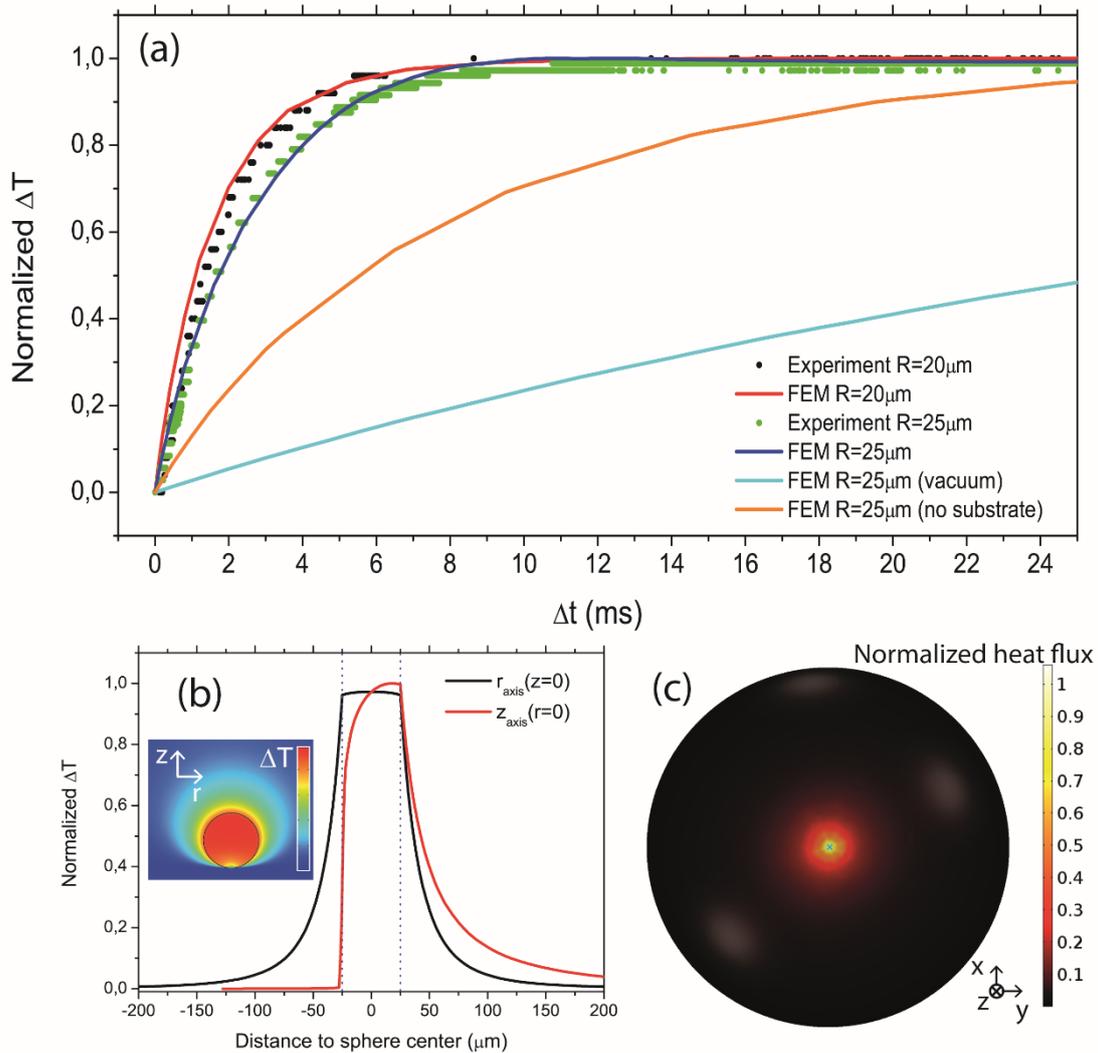

**Figure 5. (a)** Normalized experimental traces of the dynamic spectral shift (normalized temperature in the graph) obtained for microspheres of R=20 μm (black dots) and R=25 μm (green dots). Solid curves correspond to FEM simulations of the normalized thermal dynamics of the microsphere. The red (blue) curve show the case of R=20 μm (R=25 μm) surrounded by air and leaning on the substrate. The cyan (orange) curve shows the case of R=25 μm leaning on the substrate under vacuum (surrounded by moist air with no substrate). **(b)** Stationary temperature profile along the horizontal (black) and vertical (red) directions. The inset shows a 2D color scaled graph of the local temperature across a vertical plane passing through the sphere center. The profiles have been normalized to the maximum temperature within the simulation volume, which is found at the north-pole of the microsphere **(c)** Bottom view of the stationary normalized heat flux across the microsphere surface (R=25 μm). The contact point is highlighted with a cyan cross.

In conclusion, we have performed an experimental characterization of the thermal dynamics of Nd-doped BTS laser microspheres deposited on a glass substrate when they are homogeneously heated up by a pump laser. We overcome the stringent requirements for characterizing passive WGM cavities, i.e., near-field excitation of the optical modes, by using an active cavity, which allows far-field pumping and detection. By tracing the dynamic spectral behavior of the lasing modes, the heating characteristic lifetimes were determined to be about 2 ms. Those were dominated by the heat exchange towards the substrate through the medium surrounding the contact point.

Provided that the heat transfer occurs on distance scales on the hundreds of nanometers, we believe that this system and technique could become an interesting tool for studying thermal transport dynamics in non-continuum regimes. For the same reason, it would be possible to shine some light in the yet not fully understood enhancement of the radiative heat transfer [17] by performing a similar characterization below $10^{-3}$ mb, in which it becomes dominant over thermal conduction [16].

**Methods**

For the optical measurements, a continuous wave laser diode at 808 nm is modulated at a frequency of 11 Hz through a mechanical chopper. After spatial filtering by means of a pinhole it is shined into an infrared objective (Mitutoyo 20X, NA = 0.4). The resulting focused beam has a spot with an area of $1.5 \times 10^{-5}$ cm$^2$. A single microsphere can be aligned with the pump beam using a xyz micro-precision stage, the spot size being large enough to guarantee homogeneous spatial pumping over the microsphere volume. The PL is collected by an objective placed in front of the microsphere. A long-pass filter with a cut-off wavelength of 950 nm is used to filter out the laser beam excitation. The output beam is directed towards a second pinhole that removes the scattered signal. The diameter of the microspheres is far greater than the spatial resolution of the collection system (about 5 µm) [20], hence the collected PL emission comes out of localized volumes within the microsphere. In the current work we align the collection objective focal plane with the center of the microsphere and collect the signal from a region around one of its lateral edges, where the signal associated to WGMs is well-defined and not masked by the Nd$^{3+}$ bulk emission [20]. Moreover, those are the modes experiencing less radiative losses due to the presence of the substrate. Collected light is then focused at the entrance slit of a 750 mm monochromator. For signal detection, a CCD is placed at one of the monochromator output ports and directly connected to a computer. An infrared PMT (H10330-25) is placed at the other output port of the monochromator, and used in time-resolved measurements at various wavelengths. In the latter case, the signal is monitored by a digital oscilloscope using the TTL signal of the chopper as a reference. The best attainable temporal resolution is roughly 1 µs, and the maximum spectral resolution is 0.01 nm. The dimensions of the microspheres have been quantified using the collection objective and forming the device image in the CCD. Using a structure of known size as calibration, we achieved a spatial resolution of about 1 µm [20].

The microspheres were fabricated from BTS glass doped with Nd$^{3+}$ ions with the composition of 40%BaO–20%TiO$_2$–40%SiO$_2$ and doped with 1.5% Nd$_2$O$_3$ (in the molar ratio). The glass is reduced to dust by means of a mortar and is heated up to its fusion temperature, which is around 900°C. Most of the splinters melt and, when the temperature decreases, solidify in a spherical shape of several micrometer radii [14].

**Acknowledgements**

This work was supported by the EU through the Advanced Grant SOULMAN (ERC-FP7-321122) and the Spanish Ministry of Economy and Competitiveness through the project LEOMIS (TEC2012-38540-C02-01). JMR acknowledges the financial support of Secretariat for Universities and Research of Generalitat de Catalunya through the program FI-DGR 2013. We acknowledge A. Toncelli for fruitful discussions.

**Contributions**

J.M.R, D. N-U. and Y.B. performed the experiments and analyzed the data. J.M.R, D.N-U., N.E.C. and A.P. developed the model and performed the simulations. N.E.C. fabricated the microspheres. D.N-U., B.G. and A.T designed the experiment. All authors contributed to the discussion and writing of the manuscript.

**Additional Information**

The authors declare no competing financial interests.